\documentclass[seceq,supplement]{ptptex}

\usepackage{graphicx} 
\usepackage{wrapft} 
\usepackage{bbm} 

\newcommand{\be}{\begin{equation}}
\newcommand{\ee}{\end{equation}} 
\newcommand{\ba}{\begin{array}}
\newcommand{\ea}{\end{array}} 
\newcommand{\bea}{\begin{eqnarray}}
\newcommand{\eea}{\end{eqnarray}}

\newcommand{\One}{\mathbbm{1}} 
\newcommand{\lt}{\left}
\newcommand{\rt}{\right}

\newcommand{\chibar}{{\overline\chi}} 
\newcommand{\qbar}{\overline q}
 
\newcommand{\lesim}{${\lower
2pt\hbox{$\scriptstyle <$}\atop\raise 4pt\hbox{$\scriptstyle\sim$}}$}
\newcommand{\grsim}{${\lower2pt\hbox{$\scriptstyle >$}
\atop\raise4pt\hbox {$\scriptstyle\sim$}}$}

\newcommand{\naive}{na{\"\i}ve }

\newcommand{\role}{r\^ole} %
 
\title{A BCS Gapped Superfluid on the Lattice}

\author{
David N. \textsc{Walters}\footnote{This work was completed  
with the support of PPARC and in collaboration with
Simon Hands. Thanks also go to Tom Kingaby for enlightening discussions.}%
}

\inst{
Theoretical Physics Group, Department of Physics and Astronomy,
\footnote{Current address.}\\ 
University of Manchester, Manchester M13 9PL, UK\\and\\
Department of Physics, University of Wales Swansea,\\
Singleton Park, Swansea SA2 8PP, UK}

\abst{
We
discuss results of numerical analyses of the 3+1 dimensional
Nambu--Jona-Lasinio model at non-zero baryon density.
By studying relevant order
parameters, it is shown that the model goes through a transition
between a vacuum with broken chiral symmetry, and a chirally restored
phase at high baryon chemical potential $\mu$ with a diquark
condensate roughly proportional to the area of the Fermi surface and a
broken $U(1)$ baryon number symmetry.  The condensation
of diquark pairs from the fermionic spectrum is shown to lead to an
energy gap $\Delta$ about the Fermi energy $E_F$, 
which is found to be $\approx15\%$ of the vacuum fermion
mass and roughly independent of $\mu$ in the chirally restored phase.
The ground-state is believed, therefore, to be that of a BCS
superfluid. We also discuss the effect of simulating on a finite
volume, and any caveats this places on the conclusions made above.}

\begin{document}

\maketitle

\section{Introduction}

The physics of compact stars is a field that has attracted much
interest in recent years. At zero temperature, deconfined quark matter
within the core of these objects is believed to be unstable with
respect to a ground-state in which quark (and hole) pairs in the
region of the Fermi surface condense out of the spectrum in a manner
analogous to that in the BCS theory of superconductivity; because
diquark pairs cannot be colour singlet, this phenomenon is known as
colour superconductivity (CSC).\cite{Rajagopal}  The simplest CSC ground-state, known
as 2SC, is one in which condensation occurs between the approximately
degenerate $u$ and $d$ quarks, leaving an energy gap $2\Delta$ about
the Fermi surface in their spectra.  The fact that the force between
quarks in QCD is strongly attractive means that estimates for the
magnitude of $\Delta$ can be as large as 50-100MeV, i.e. large enough
to remain present at typical compact star temperatures of 
${\cal O}$(1MeV).\cite{Berges}

The confirmation of this picture by a first principles lattice QCD
calculation, however, remains elusive due to the difficulties of
performing Monte-Carlo simulations with baryon chemical potential
$\mu\neq0$. Whilst the ``sign problem'' remains, the only way of
probing the cold, dense region of the QCD phase diagram
non-perturbatively is by using model field theories, such as 2 colour
QCD or the Nambu--Jona-Lasinio (NJL) model.

In both of these models, the functional weight remains both real and
positive, even with $\mu\neq0$,\cite{2CQCD,NJL3} 
such that they are simulable using
standard lattice techniques. Lattice simulations of the NJL model 
show that chiral
symmetry restoration occurs at some critical chemical potential
$\mu_c\sim\Sigma_0$, where $\Sigma_0$ is the constituent quark mass,
in qualitative agreement with analytic approaches such as the
large-$N_c$ (or Hartree) approximation.\cite{Klevansky}  
Furthermore, whilst this model has no gauge degrees
of freedom and cannot describe confined matter, the fundamental quarks that
play the \role~of the baryons do obey the correct statistics, such
that the effect of increasing $\mu$ beyond $\mu_c$ is to build up a Fermi
surface. This should be contrasted with 2 colour QCD, in which the $q
q$ baryons are bosonic and the onset of matter occurs at $\mu_c\sim
m_\pi/2$. Together, these features make the lattice NJL model an ideal tool
with which to study Fermi surface phenomena, such as CSC, at
phenomenologically relevant densities.

\section{The model} \label{sec:themodel}

The model studied herein is the 3+1 dimensional lattice NJL model,
which in continuum notation has the Euclidean Lagrangian 
\bea 
{\cal L}&=&\bar\psi(\partial{\!\!\!/\,}+m+\mu\gamma_0)\psi\nonumber\\
&-&{\tfrac{g^2}{2}}\Bigl[(\bar\psi\psi)^2
-(\bar\psi\gamma_5\otimes\vec\tau\otimes\gamma_5\psi)^2\Bigr]\label{eq:L}\\
&+&{\tfrac{1}{2}} \Bigl[ (\bar\psi,\psi^{t r})\left(\ba{c c} j& \\
&j\ea\right) C\gamma_5\otimes\tau_2\otimes C\gamma_5
{\scriptstyle{\left(\ba{c}\bar\psi^{t r}\\\psi\ea\right)}}\Bigr],
\nonumber 
\eea 
where $m$ is the bare- or current-quark mass, and the
source $j$ has been added to allow the measurement of a baryon number
violating diquark condensate in a finite volume system.

When formulated on a discrete lattice of sites $x$ with separation $a$
between nearest neighbours, the model is described in terms of the
staggered fermion fields $\chi$, $\chibar$, $\zeta$ and
$\overline\zeta$ by the action 
\be 
S=a^4\sum_x\lt[\lt(\chibar,\chi^{t r}\rt){\cal A}\lt(\ba{c}\chibar^{t
r}\\\chi\ea\rt)+\lt(\overline\zeta,\zeta^{t r}\rt){\cal
A}^*\lt(\ba{c}\overline\zeta^{t r}\\\zeta\ea\rt)\rt]
+\frac{2a^4}{g^2}\sum_{\tilde x}\lt(\sigma^2+\vec\pi.\vec\pi\rt),
\label{eq:S} 
\ee 
where ${\cal A}$ is the Nambu-Gor'kov matrix 
\be
{\cal A}=\frac{1}{2}\lt(\ba{cc}j\tau_2^{p q}&M\\ -M^{t r}&j\tau_2^{p
q}\ea\rt).  
\ee 
In turn, $M$ is the fermion kinetic matrix 
\bea M^{p q}_{x y} & = & \frac{1}{2a}\delta^{p q} \lt[\lt(e^{a\mu}\delta_{y
x+\hat{0}}-e^{-a\mu}\delta_{y x-\hat{0}}\rt) + \sum_{\nu=1}^3
\eta_\nu(x)\lt(\delta_{y x+\hat{\nu}} -\delta_{y x-\hat{\nu}}\rt)
+2am\delta_{x y}\rt] \nonumber \\ & + & \frac{1}{16}\delta_{x
y}\sum_{\lt<\tilde{x},x\rt>} \lt(\sigma(\tilde{x})\delta^{p q} +
i\epsilon(x)\vec{\pi} (\tilde{x}).\vec{\tau}^{p q}\rt),
\label{eq:fullM} 
\eea 
where $\vec\tau^{p q}$ is a vector of the Pauli
matrices which act on the $N_f=2$ isospin degrees of freedom. The
lattice model also exhibits an additional $N_c=8$ ``colour'' degrees
of freedom, due to staggered fermion doubling and the choice of
(\ref{eq:S}).  The auxiliary scalar and pseudoscalar fields $\sigma$
and $\vec\pi$ are defined on sites $\tilde x$ of a dual lattice,
$\lt<\tilde x,x\rt>$ represent the sum over the 16 dual sites neighbouring
$x$, and the symbols $\eta_\nu(x)$ and $\epsilon(x)$ are the phases
$(-1)^{x_0+\dots+x_{\nu-1}}$ and $(-1)^{x_0+x_1+x_2+x_3}$
respectively.

In the limit that $m$ and $j\to0$, (\ref{eq:S}) has the same global
symmetries as QCD with two massless flavours, i.e. the $SU(2)_L\otimes
SU(2)_R$ chiral symmetry of left- and right-handed quarks 
\be
\ba{cc}\ba{cc}
\chi\to(P_L U+P_R V)\chi;&\chibar\to\chibar(P_L V^\dagger+P_R U^\dagger);\ea 
\vspace{0.5ex}\\\ba{cc}
\zeta\to(P_L V+P_R U)\zeta;&\overline\zeta\to
\overline\zeta(P_L U^\dagger+P_R V^\dagger);\ea
\vspace{0.5ex}
\\(\sigma+i\vec\pi.\vec\tau)\to V(\sigma+i\vec\pi.\vec\tau)U^\dagger,
\ea\ee 
where
$P_{L,R}\equiv\tfrac{1}{2}\lt(1\pm\epsilon(x)\rt)$ and $U$ and $V$ are
elements of $SU(2)$, and the $U(1)_B$ symmetry of baryon number
\be\ba{cc} 
\chi\to e^{i\alpha}\chi,&\chibar\to e^{-i\alpha}\chibar.
\ea\ee 
As in QCD, at $\mu=0$ chiral condensation leads the spontaneous
generation of a constituent quark mass $\Sigma_0\gg m$ which breaks
$SU(2)_L\otimes SU(2)_R$ down to the $SU(2)_I$ of isospin. These
reassuring features suggest that by studying the model's global
symmetries and their breaking, we can learn about the analogous
symmetry breaking patterns in QCD.

In 3+1 dimensions the NJL model is only an effective theory, such that
the lattice parameters must be chosen to match low energy
phenomenology. By applying the large-$N_c$ (Hartree) approximation
analytically to staggered quarks, we determine a suitable set, leading
to physically reasonable results, to be 
\begin{equation}
\begin{array}{c c c} 
\begin{array}{r c l}
am&=&0.006\\a^{2}g^{-2}&=&0.495\\a^{-1}&=&720\mbox{\rm MeV}
\end{array}&\Rightarrow& \begin{array}{r c l}
\Sigma_0&=&400\mbox{\rm MeV}\\f_\pi&=&93\mbox{\rm
MeV}\\m_\pi&=&138\mbox{\rm MeV}.
\end{array}\end{array} \label{eq:contnos} \end{equation}

\section{Order parameters}
\label{sec:orderparam}

\begin{figure}[ht]
\centering
\includegraphics[width=10cm]{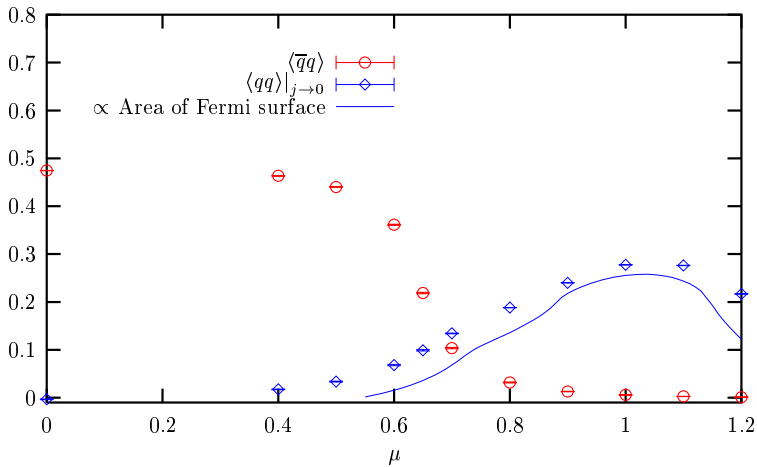} 
\caption{Order parameters as
functions of $\mu$. The solid line represents the area of the
Fermi surface calculated in the large-$N_c$ limit.}  
\label{fig:EofS} 
\end{figure}

In previous work\cite{NJL4v1}, we have performed simulations of the model outlined
in \textsection\ref{sec:themodel} in an attempt to identify diquark
condensation in the dense phase. In particular, we measure the diquark
condensate 
\be 
\lt<q q\rt>\equiv\frac{1}{2V}\frac{\partial\ln{\cal
Z}}{\partial j} =\frac{1}{8V}\lt<{\rm Tr}\lt(\ba{c
c}\tau_2&\\&\tau_2\ea\rt){\cal A}^{-1}\rt>, 
\ee 
where $\cal Z$ is the partition function, 
and compare it to the chiral condensate 
\be \lt<\qbar q\rt>\equiv\frac{1}{V}\frac{\partial\ln{\cal Z}}{\partial m}
=\frac{1}{4V}\lt<{\rm Tr}\lt(\ba{c c}&\One_2\\-\One_2& \ea\rt){\cal
A}^{-1}\rt>.  
\ee 
We perform simulations for a range of $\mu$ on
$V=L_t\times L_s^3$ lattices with $L_t=L_s=12$, 16 and 20 using a
standard hybrid Monte-Carlo (HMC) algorithm, in which the source
strength is set to zero during the update of the auxiliary background
fields, but allowed to vary over a range of $j=j^*\neq0$ during the
quenched measurement of the quantities above. An interesting empirical
observation is that results in the chiral sector are found to scale
linearly with inverse volume, whilst diquark observables scale with inverse
temporal extent, which in Euclidean space corresponds to the
temperature.  Accordingly, $\lt<\qbar q\rt>$ is extrapolated
to $V^{-1}\to0$, whilst $\lt<q q\rt>$ is extrapolated first to
$L_t^{-1}\to0$ and then $j\in[0.3,1.0]\to0$.\footnote{Although
all diquark observables are measured for
$0.1\leq j\leq1.0$, 
the behaviour of data with $j<0.3$ appears to
disagree markedly with that of those at higher $j$, and these are ignored for the purposes of
these fits. We attribute this to finite
volume errors, which we discuss briefly in \textsection\ref{sec:finitev}.} 
The results are plotted in Fig.~\ref{fig:EofS}.  
As expected, at $\mu=0$
chiral symmetry is broken leading to a non-zero chiral condensate
whilst $\lt<q q\rt>\simeq0$. As chemical potential is increased, the
system goes through a crossover into a phase with approximate chiral
symmetry restoration and a non-zero diquark condensate. In particular,
$\lt<q q\rt>$ is found to increase roughly as the area of the lattice
Fermi surface, plotted in arbitrary units for free fermions in the
large-$N_c$ limit as the solid curve. 
This supports the \naive picture of the condensate being a
measure of the density of pair states about the Fermi surface,
contributing to diquark condensation.

\section{The superfluid gap}
\label{sec:gap}

In the previous section, we presented evidence for superfluidity in
the form a non-zero diquark condensate in the high-$\mu$ phase. This
evidence must be treated as indirect, however, since this quantity is
not measurable in an experiment. Furthermore, even if the sign problem
were overcome, one could not measure $\lt<q q\rt>$ in lattice QCD,
since in this theory the colour superconducting phase is
characterised by the breaking of a gauge symmetry; Elitzur's theorem
states that one cannot define a local order parameter to distinguish
the existence of such a phase in a gauge invariant way.\cite{Elitzur} 
Instead, we
turn to more direct evidence in the form of the global order
parameter for the BCS phase, the energy gap $\Delta$.

We do this by studying the fermionic dispersion relation $E(k)$,
extracted from the Gor'kov time-slice propagator 
\be 
{\cal G}(\vec k,t)\equiv\sum_{\vec x}{\cal A}^{-1}(\vec0,0;\vec x,t) 
e^{-i\vec k.\vec x}, 
\ee 
on $L_t\times L_x\times L_{y,z}^2$ lattices with
$L_t=16$, 20 and 24, $L_x=96$ and $L_{y,z}=12$. The momenta sampled
are $\vec k=(2\pi n/L_x,0,0)$ for $n=0$, 1, 2, $\ldots$, implying
that for our choice of lattices there are 25 independent momentum
modes between $k=0$ and $\pi/2$ in the $k_x$ direction.  We measure
${\cal G}$ at $\mu=0.8$ for a range of $j$, using standard lattice
methods for meson correlators, in the same partially
quenched approximation employed in the measurement of the order
parameter. It is found empirically that whilst in general $\cal G$
has 16 complex components in Nambu-Gor'kov space, only two parts of these
are independent and non-zero. These are 
\bea 
N(k,t)\equiv{\rm Re}({\cal G}_{1 3}(k,t)) &~~~
\&~~~& A(k,t)\equiv{\rm I m}({\cal G}_{2 1}(k,t)), 
\eea 
which in the limit that $j\to0$ correspond to the
propagators for particles 
\be 
N(k,t)\sim\lt<q(x)\bar q(y)\rt>_{11} 
\ee
and for superpositions of particles and holes 
\be
A(k,t)\sim\lt<q(x)q(y)\rt>_{21}.  
\ee 
We label these the ``normal''
and ``anomalous'' propagators respectively. Fitting these to 
\begin{subequations}
\be
N(k,t)=A e^{-E t}+Be^{-E(L_t-t)} 
\label{eq:Nfit} 
\ee
\be
A(k,t)=C(e^{-E t}-e^{E(L_t-t)}) 
\label{eq:Afit} 
\ee
\end{subequations}
\begin{wrapfigure}{r}{6.6cm}
\includegraphics[width=6.6cm,height=5cm]{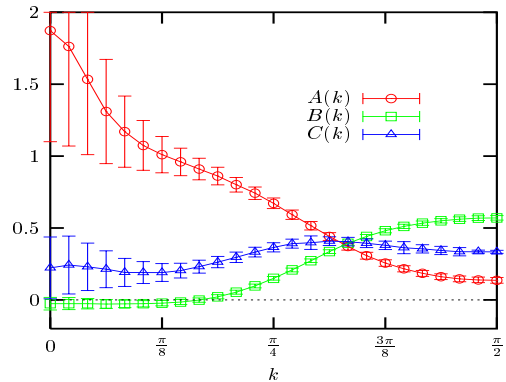}
\caption{Propagator amplitudes at $\mu=0.8$.}  
\label{fig:ABCvsk} 
\end{wrapfigure}
we extract the amplitudes $A(k,j)$, $B(k,j)$
and $C(k,j)$ and the excitation energy $E(k,j)$, and as
with $\lt<q q\rt>$ extrapolate to $L_t^{-1}\to0$ and then
$j\in[0.3,1.0]\to0$.

Fig.~\ref{fig:ABCvsk} shows the amplitudes $A$, $B$ and $C$ as
functions of $j$ in the limit $T,j\to0$. At low $k$, $A\gg B$
corresponding to a predominantly forward moving signal, which implies
that the normal propagator deep within the Fermi sea is dominated by
hole-like excitations. At large $k$ the reverse is true, implying that
excitations above the Fermi surface are dominated by particles. The
point at which $A(k)$ and $B(k)$ coincide, therefore, can be
identified with the momentum at the Fermi surface, $k_F$.\cite{NJL3} 
Of particular
interest is the fact that the coefficient for the anomalous propagator
$C(k)$ is non-zero in a broad peak about $k_F$, even in the $j\to0$
limit, which implies that there are particle-hole mixed states with indefinite
baryon number, an indirect signal for superfluidity via a BCS
mechanism.

\begin{figure}[ht] 
\centering
\includegraphics[height=5.3cm]{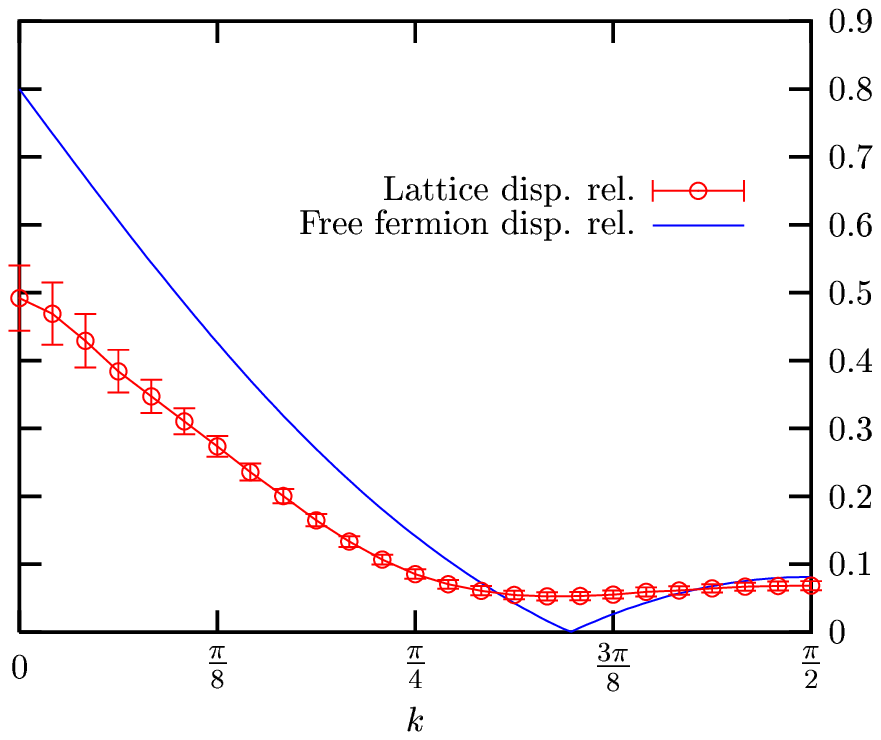}
\includegraphics[height=5.3cm]{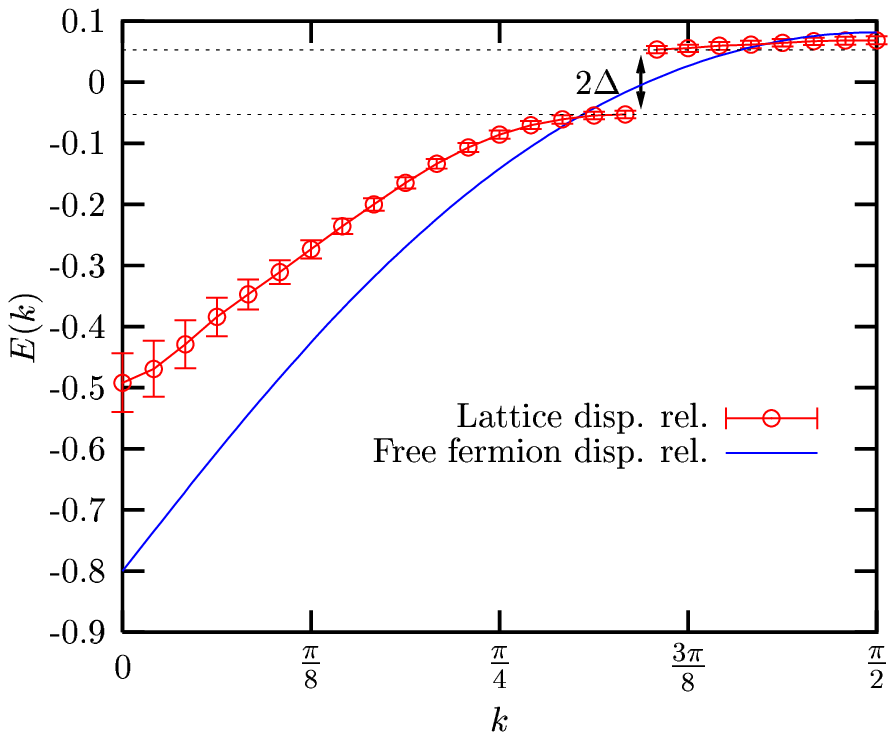} 
\caption{Lattice dispersion relation and typical free fermion
dispersion relation at $\mu=0.8$.}  
\label{fig:mu0.8Evsk} 
\end{figure} 
The left-hand panel of
Fig.~\ref{fig:mu0.8Evsk} shows the fermion dispersion relation at $\mu=0.8$,
i.e. the energy of the fermionic excitations as a function of their
momenta, compared to that of non-interacting lattice
fermions. The free fermion dispersion relation clearly has two
branches, representing hole excitations for $k<k_F$ and particle
excitations for $k>k_F$. In between one observes that
$E(\sin^{-1}(\sinh(\mu a)))=0$. In contrast to this, in the dispersion
relation extracted from the lattice data one can see no discontinuity
between the two branches, another sign of particle-hole mixing. More
importantly, at no point does $E(k)$ pass through $E=0$ and there is a
distinct gap between this point and the minimum.

This can be seen in a more familiar light if one plots the hole
branch as negative, which is done in the right-hand panel of
Fig.~\ref{fig:mu0.8Evsk}. This makes the free fermion dispersion
relation a smooth continuous curve, which is similar to those observed
in lattice theories with no BCS gap,\cite{NJL3,Hands} whilst for our
smooth dispersion relation
this introduces a discontinuity at $k\approx k_F$ signalling an energy
gap $2\Delta$. This is the first direct
observation of a BCS gap in a lattice simulation.

In order to learn something of the chemical potential
dependence of the gap we perform a \naive extrapolation of data
extracted from $L_t=16$ and 20 lattices to zero
temperature and assign a conservative estimate of the error; the
$L_t=24$ data required for a full statistical treatment such as that
applied at $\mu=0.8$ are too expensive to reproduce for a range of $\mu$
in the chirally restored phase. Data are then extrapolated to $j\to0$
as before so that we can plot the dispersion relations and gain 
an estimate for the gap.  
\begin{figure}[ht]
\centering 
\includegraphics[height=6cm]{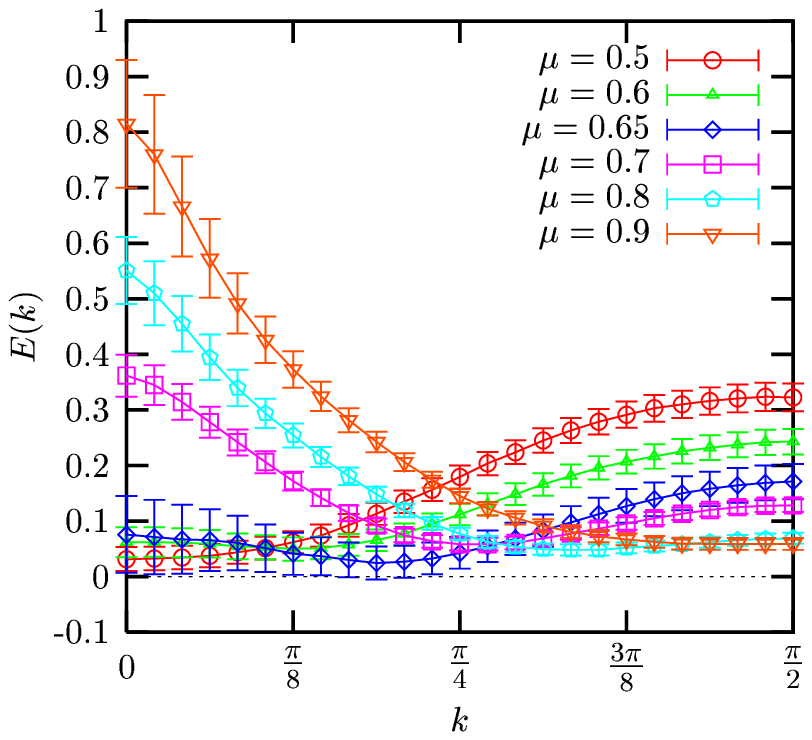} 
\includegraphics[height=6cm]{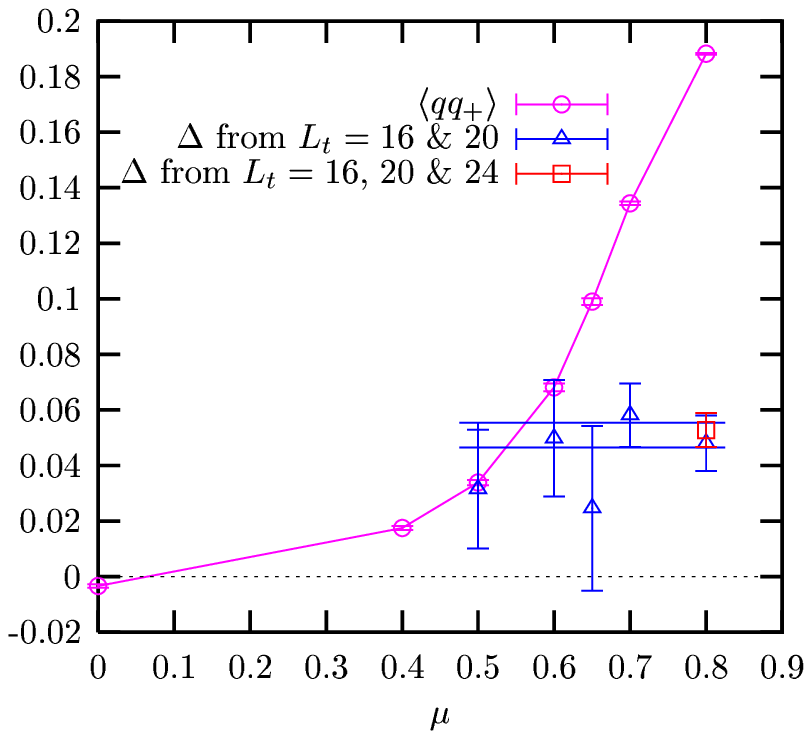} 
\caption{Lattice dispersion relations at various values of $\mu$ and
a comparison between their minima and the diquark condensate.}
\label{fig:Evsk} 
\end{figure}
These are plotted in the left-hand panel of 
Fig.~\ref{fig:Evsk}. It is
worth noting that whilst a linear extrapolation through two points is
of little statistical value, the results produced at $\mu=0.8$ with and without
using the $L_t=24$ data are found to be consistent for all $k$.

The estimates for $\Delta(\mu)$ are plotted in the right-hand panel of
Fig.~\ref{fig:Evsk}, which shows that 
$\Delta$ appears roughly constant for $\mu>\mu_c$, whilst $\lt<q
q\rt>$ rises with the area of the Fermi surface. This is consistent
with the simple-minded assumption that the order parameter counts the
density of states from within a shell of thickness ${\cal O}(\Delta)$
about the Fermi surface that contribute to diquark condensation, such that
in the continuum
\be
\lt<q q\rt>\propto\Delta\mu^2.
\ee
Finally, in order to express our best measurement of the
gap, $\Delta a(\mu a=0.8)=0.053(6)$, in a form independent of the lattice
spacing $a$, we extract the vacuum fermion mass $\Sigma_0$ from
simulations with 
$L_t=16$, 20 and 24 and find
\be
\frac\Delta{\Sigma_0}=0.15(2).
\ee
Assuming that $\Sigma_0\sim400$MeV, this implies that
in physical units, the size of the gap is $\Delta\sim60$MeV,
which is consistent with the predictions of Ref.~\citen{Berges}.

\section{Finite volume effects}
\label{sec:finitev}

An important point that we have failed, thus far, to highlight is that
the conclusions of the previous sections rely on the disregarding of
data with $j<0.3$, for which the order parameter
and other diquark observables deviate sharply from the fitted form. We
attribute this to finite volume errors, which are 
particularly large in this system,\cite{Amore} a fact we
believe to be due to the difficulty of representing a small shell of
states close to the Fermi surface on a discrete momentum-mode lattice.
To address this issue, parallel simulations are currently being attempted on
very large lattices, with the aim of showing that the
on such systems the deficiency at low-$j$ is reduced. 
Thusfar, these simulations have proven hard, with very small time-steps
and trajectories required for acceptance rates of over 50\%. 
This might be interpreted as a
sign of encouragement, however, as the failure of the algorithm may be a signal
that the partially quenched approximation is insufficient for such
large volumes. If the volume is large enough that the proposed
finite size effects are overcome, there may be enough momentum
modes within the shell about the Fermi surface for diquark
condensation to occur  even with
$j\equiv0$, as partial quenching enforces during the field updates. The
algorithm would then be attempting to represent a system with an exact
Goldstone mode and encounter the standard finite size effects related
to divergent Goldstone fluctuations. 
A full simulation, with $j\neq0$ during all steps of
the algorithm, may yet be required to resolve this issue.

\section{Summary}

To summarise, our evidence for s-wave superfluidity via a BCS
instability is of some importance, since this is the first time
that the presence of such a phase has been demonstrated in a
relativistic quantum field theory using a systematic calculational
technique. Although the $3+1d$ NJL model is only a simplistic
effective field theory, this work  can be interpreted
phenomenologically as non-perturbative evidence for  BCS colour
superconductivity in QCD with two degenerate flavours, since these two
theories have the same global symmetry structure. This is of
particular importance whilst the persistence of the sign problem
prevents the numerical solution of full $SU(3)$ QCD. 
Also, the fact that our 
results agree with the predictions of self-consistent
treatments of this simple model adds credence to solutions of
similar models with more complicated flavour structures and 
interactions.

In future work it would be interesting to further this analysis to study the
stability of the superfluid phase by performing simulations either at non-zero
temperature with the aim of measuring the critical temperature of the
superfluid phase, or with the Fermi surfaces for ``up'' and ``down''
quarks separated via the introduction of a small non-zero isospin chemical
potential. The latter option may be difficult, however, since initial
studies indicate that with this introduction, $\det M$ becomes
complex. 

The most pressing issue, however, is to resolve the
finite volume effects discussed above, since it is only once this
issue has been addressed that the conclusions of
\textsection\ref{sec:orderparam} \& \textsection\ref{sec:gap} may be
trusted completely. 


%

 \end{document}